\documentclass[%
 reprint,
superscriptaddress,
 amsmath,amssymb,
 aps,
 pra,
showkeys]{revtex4-1}

\usepackage{soul,xcolor} 
 
\usepackage{tcolorbox}
\usepackage{graphicx}
\usepackage{dcolumn}
\usepackage{bm}
\usepackage{amsthm}
\usepackage{hyperref}
\hypersetup{
    colorlinks=true,
    linkcolor=blue,
    filecolor=magenta,      
    urlcolor=cyan,
}

\usepackage{natbib}
\usepackage{color}

\usepackage{siunitx}

\newtheorem*{theorem}{Problem}

\usepackage{xpatch}
\makeatletter
\AtBeginDocument{\xpatchcmd{\@thm}{\thm@headpunct{.}}{\thm@headpunct{}}{}{}}
\makeatother

\begin{document}

\title{Machine learning techniques for state recognition and auto-tuning in quantum dots}

\author{Sandesh S. Kalantre}
\email{sandeshkalantre@iitb.ac.in}
\affiliation{Department of Physics, Indian Institute of Technology - Bombay, Mumbai, 400076, India}
\affiliation{Joint Center for Quantum Information and Computer Science, 
University of Maryland, College Park, MD, 20742, USA}
\affiliation{National Institute of Standards and Technology, Gaithersburg, MD, 20899, USA}

\author{Justyna P. Zwolak}
\affiliation{Joint Center for Quantum Information and Computer Science, 
University of Maryland, College Park, MD, 20742, USA}
\affiliation{National Institute of Standards and Technology, Gaithersburg, MD, 20899, USA}

\author{Stephen Ragole}
\affiliation{Joint Center for Quantum Information and Computer Science,
University of Maryland, College Park, MD, 20742, USA}
\affiliation{Joint Quantum Institute, University of Maryland, College Park, MD, 20742, USA}

\author{Xingyao Wu}
\affiliation{Joint Center for Quantum Information and Computer Science,
University of Maryland, College Park, MD, 20742, USA}
\affiliation{Joint Quantum Institute, University of Maryland, College Park, MD, 20742, USA}

\author{Neil M. Zimmerman}
\affiliation{National Institute of Standards and Technology, Gaithersburg, MD, 20899, USA}

\author{M. D. Stewart, Jr.}
\affiliation{National Institute of Standards and Technology, Gaithersburg, MD, 20899, USA}

\author{Jacob M. Taylor}
\email{jmtaylor@umd.edu}
\affiliation{Joint Center for Quantum Information and Computer Science,
University of Maryland, College Park, MD, 20742, USA}
\affiliation{National Institute of Standards and Technology, Gaithersburg, MD, 20899, USA}
\affiliation{Joint Quantum Institute, University of Maryland, College Park, MD, 20742, USA}
\affiliation{Research Center for Advanced Science and Technology,
University of Tokyo, Meguro-ku, Tokyo 153-8904, Japan}

\date{\today}
 \keywords{semiconductor quantum computation; quantum dots; machine learning; convolutional neural networks; auto-tuning}
\begin{abstract}
Recent progress in building large-scale quantum devices for exploring quantum computing and simulation paradigms has relied upon effective tools for achieving and maintaining good experimental parameters, i.e., tuning up devices. In many cases, including in quantum-dot based architectures, the parameter space grows substantially with the number of qubits, and may become a limit to scalability. Fortunately, machine learning techniques for pattern recognition and image classification using so-called deep neural networks have shown surprising successes for computer-aided understanding of complex systems. 
In this work, we use deep and convolutional neural networks to characterize states and charge configurations of semiconductor quantum dot arrays when one can only measure a current-voltage characteristic of transport through such a device. 
For simplicity, we model a semiconductor nanowire connected to leads and capacitively coupled to depletion gates using the Thomas-Fermi approximation and Coulomb blockade physics. 
We then generate labeled training data for the neural networks, and find at least $90\,\%$ accuracy for charge and state identification for single and double dots purely from the dependence of the nanowire's conductance upon gate voltages. 
Using these characterization networks, we can then optimize the parameter space to achieve a desired configuration of the array, a technique we call `auto-tuning'.
Finally, we show how such techniques can be implemented in an experimental setting by applying our approach to an experimental data set, and outline further problems in this domain, from using charge sensing data to extensions to full one and two-dimensional arrays, that can be tackled with machine learning.
\end{abstract}
\maketitle

\section{Introduction}
Tremendous progress in realizing high-quality quantum bits at the few qubit level has opened a window for new challenges in quantum computing: developing the necessary classical control techniques to scale systems to larger sizes. A variety of approaches~\cite{Li,Karzig2017,Neill2017,Zajac,Saffman2016,Sete2016,Blais2004,Brown2016,Bernien17} rely upon tuning individual quantum bits into the proper regime of operation. In semiconductor quantum computing, devices now have tens of individual electrostatic and dynamical gate~\cite{Fogarty2017,Malinowski2017,Reed2016,Jones2016,Veldhorst2016,Nichol2016,Zajac,Delbecq2014,Tosi2017} voltages which must be carefully set to isolate the system to the single electron regime and to realize good qubit performance. A similar problem arises in the control of ion positions in segmented ion traps~\cite{Bermudez2017,Furst2014,Alonso2013,Schulz2008}. Preliminary work to automate the laborious task of tuning such systems has primarily focused on fine tuning of analog parameters~\cite{Watson2017,Baart2016} using techniques from regression analysis and quantum control theory. At the same time, tremendous progress in automated classification suggests such techniques may be used~\cite{Szegedy2015,AlexNet,LeCun1998} to bootstrap the experimental effort from a \emph{de novo} device to a fully tuned device, replacing the gross-scale heuristics, developed by experimentalists to deal with tuning of parameters particular to experiments.


In this work, we specifically consider the control problems associated with electrostatically defined quantum dots (QDs) present at the interface of semiconductor devices~\cite{Wiel2003}. Each quantum dot is defined using voltages applied to metallic gate electrodes acting as depletion gates which confine a discrete number of electrons to a set of \emph{islands}. We use machine learning (ML) and numerical optimization techniques to efficiently explore the multidimensional gate voltage space to find a desired island configuration, a technique we call `auto-tuning'. Toward this end, we use ML to recognize the number of dots generated in the experiment. 

In order to improve on the accuracy, we work with convolutional neural networks (CNNs)~\cite{AlexNet,LeCun15}. CNNs  are a class of artificial neural networks designed for efficient pattern recognition and classification of images. When trained on high quality simulated data, CNNs can learn to identify the number of QDs. Once the neural network is trained to recognize dot configurations, we can recast the problem of finding a required configuration as an optimization problem. As a result, a neural network coupled to a optimization routine presents itself as a solution for determining a suitable set of gate voltages.

Training of a machine learning algorithm necessitates the existence of a physical model to qualitatively mimic experimental output and provide a large, fully labeled data set. In this paper, we develop a model for transport in gate-defined quantum dots and train neural networks to identify number of islands under a given gate voltage configuration. We also describe the auto-tuning problem in the double-dot to single dot transition regime. Finally, we discuss the performance of the recognition and auto-tuning for both simulated and experimental data. We report over $90\,\%$ accuracy for with very simple neural network architectures on all these problems, where accuracy is defined as the fraction of times when the predicted configuration agreed with the pre-assigned label. 

This paper is organized as follows. In Section~\ref{sec:motivation}, we motivate the problems associated with tuning of quantum dot arrays and their relation to ML problems. In Section~\ref{sec:physical-model}, we present the physical setup for our devices and the model for transport calculations. In Section~\ref{sec:learning-Coulomb-blockade}, we start with a toy example of using a neural network to learn Coulomb blockade and identify charge states of a single quantum dot. The charge and state identification problem for a double dot and its solution using CNNs is presented in Section~\ref{sec:learning-state}. In Section~\ref{sec:auto-tuning}, we define the auto-tuning problem and its resolution. In Section~\ref{sec:experimental}, we test our trained CNN for state identification and auto-tuning on experimental data. In Section~\ref{sec:discussion}, we describe how the machine learning techniques described in this work can be incorporated in an experimental setting and speculate on further problems that can be potentially solved using neural networks for quantum dots. Finally, we present our conclusions in Section~\ref{sec:conclusion}.

\section{Motivation}\label{sec:motivation}
\begin{figure}[b]
\centering
\includegraphics[width=\linewidth]{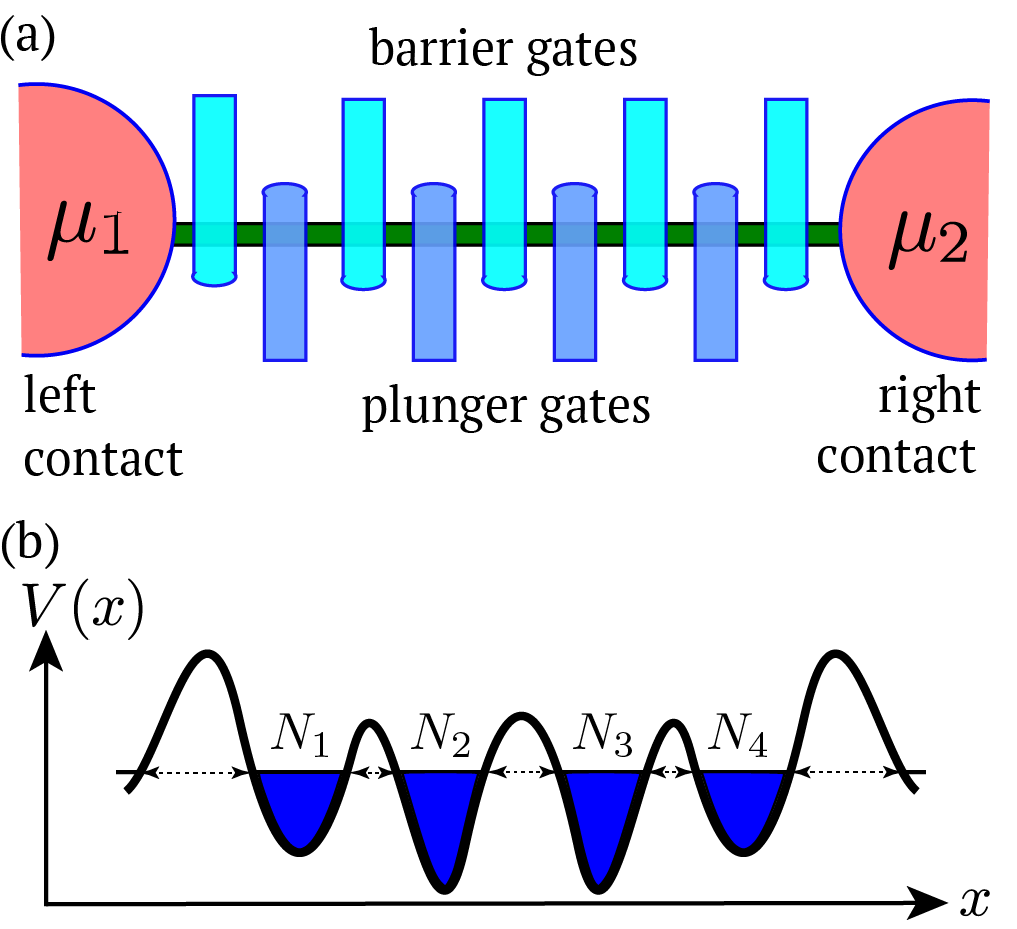}
\caption{(a) A generic nanowire connected to contacts with top gates. $\mu_1$ and $\mu_2$ are the chemical potentials of the contacts. (b) Potential profile $V(x)$ along the nanowire. Alternating set of barrier and plunger gates create a potential profile $V(x)$ along the nanowire. ($N_1,N_2,N_3$, and $N_4$ are the number of electrons on each island. Electrons can tunnel through the barriers between adjacent islands or the contacts. The filled blue areas denote regions of electron density.)} \label{fig:nanowire_model}
\end{figure}

Electrostatically defined quantum dots offer a means of localizing electrons in a solid-state environment. A generic device, consisting of a linear array of dots in a two-dimensional electron gas (2DEG), is presented in Fig.~\ref{fig:nanowire_model}(a). Gate electrodes on top are used to confine electron density to certain regions, forming islands of electrons. The ends of the linear array are connected to reservoirs of electrons, i.e., contacts, which are assumed to be kept at a fixed chemical potential. 

By applying suitable voltages to the gates, it is possible to define a one dimensional potential profile $V(x)$. Alternating regions of electron density islands and barriers are formed, depending on the relation between the chemical potential and the electrostatic potential $V(x)$ (Fig.~\ref{fig:nanowire_model}(b)). Barrier gates are used to control tunneling between the islands while the plunger gates control the depths of the potential wells.

A fixed number of islands requires a specific number of gates. Since the voltage on each gate can be set independently, the state space for the gate voltages is $\mathbb{R}^m$, with $m$ denoting the number of gates. By suitable choices of the gate voltages, it is possible to have a certain number of islands, each with a certain number of charges along the nanowire. We refer to the number of islands as the state. Though having a large number of gates implies a higher degree of control, it also presents a challenge in determining appropriate values for the gate voltages, given a required configuration \cite{Zajac}.

Standard techniques of assigning voltages to the gates rely on heuristics and experimental intuition. Such techniques, however, present practical difficulties in implementation when the number of gates increases beyond a modest number. Hence, it is desirable to have a technique, given a desired configuration of the device, to determine an appropriate voltage set without the need for actual intervention by an experimenter.

Machine learning (ML) \cite{Goodfellow-et-al-2016} is an algorithmic paradigm in artificial intelligence and computer science to learn patterns in data without explicitly programming about the characteristic features of those patterns. An important task in machine learning is classification of data into categories, generically referred to as a classification problem. The algorithm learns about the categories from a dataset and produces a model that can assign \emph{previously unseen} inputs to those categories.

In supervised learning models, ML algorithms rely on \emph{labels} identifying each data point to learn to classify data from
a predefined and known representative subset (\emph{training data}) into assumed categories (thus the term \emph{supervised}). Once trained, the algorithm then generalizes to an unknown data set, called the \emph{test set}. Deep neural networks (DNNs) i.e., neural networks with multiple hidden layers, can be used to classify complex data into categories with high accuracy of over $90\,\%$ \cite{AlexNet}.

The central aim of this work is to enable an automated approach to navigation and tuning of quantum dot devices in the multidimensional space of gate voltages. Here, we define auto-tuning specifically as finding appropriate values for the gate electrodes to achieve a particular configuration. Identification of the state of the device is the first step in the tuning process. In light of the requirement for learning the state to achieve tuning and the success achieved with DNNs for data classification, we propose to use DNNs to determine charges and states of quantum dots. Once it is achieved, auto-tuning is reduced to an optimization problem to the required state and can be done with standard optimization routines.

\section{Physical model of a nanowire}\label{sec:physical-model}
A prerequisite for training of neural networks is the availability of a training data set which mimics the expected characteristics from a test set. We develop a model for electron transport under the Thomas-Fermi approximation to calculate electron density $n(x)$ and current $I$ (see Appendix~\ref{appendix:tf} for details). This model allows us to construct a capacitance model for the islands given a potential landscape $V(x)$ and the fixed Fermi level of the contacts. The potential profile, in turn, is determined by the voltages set on the gates (Appendix~\ref{appendix:gate}).

An infinitesimal bias is assumed to exist between the contacts. The discreteness in the number of electrons in the islands, along with inter-electronic Coulomb repulsion, leads to transport being blockaded across the nanowire. The charge configuration changes when there are two or more degenerate charge states. Such a degeneracy in energy leads to electron flow across the leads, i.e., current at an infinitesimal bias.

We model electron transport using a Markov chain among the charge states $(N_1,N_2,\dots,N_k)$ of the $k$ islands. $N_i$ represents the number of electrons on the $i^{th}$ island. The rate of going from one state to another is calculated under a thermal selection rule set by the energy of the two configurations evaluated from the capacitance model and the tunneling rate. The tunneling rate is modeled as a product of the Wentzel-–Kramers-–Brillouin (WKB) tunnel probability \cite{merzbacher1998quantum} across the barrier and classical attempt rate of electrons in the islands. From the steady state configuration of the Markov chain, we calculate the current for a given potential landscape $V(x)$ (Appendix~\ref{appendix:markov}). In all, this simplistic approach provides the minimum model to reproduce basic charge configurations and transport characteristics qualitatively for linear arrays of quantum dots.

\begin{figure}[t]
\centering
\includegraphics[width=\linewidth]{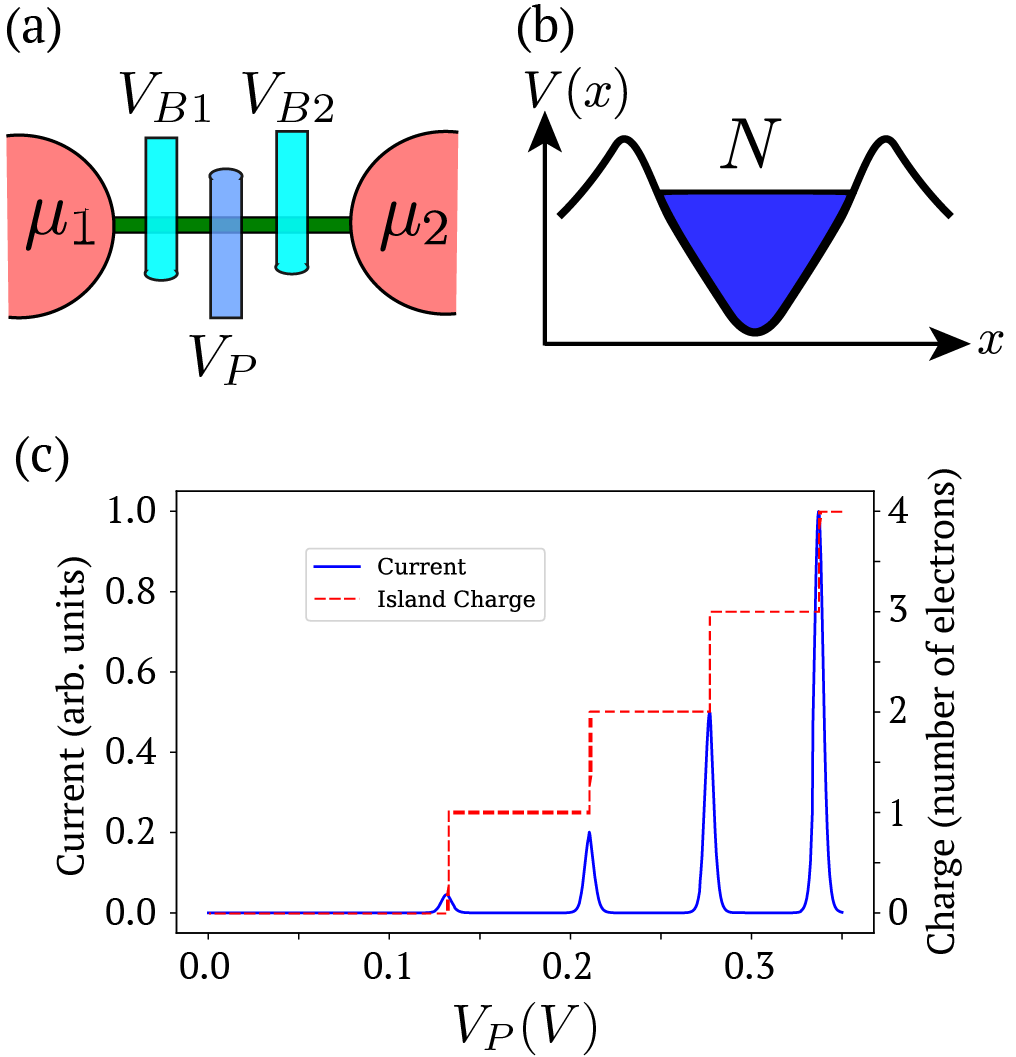}
\caption{(a) A single dot device model. (b) Potential profile $V(x)$ along the nanowire with a single dot. (c) Simulated current and electron number $N$ for a single dot exhibiting Coulomb blockade as a function of plunger gate voltage $V_P$ for a 3-gate device}
\label{fig:3-gates-model}
\end{figure}

\begin{figure}[t]
\centering
\includegraphics[width=\linewidth]{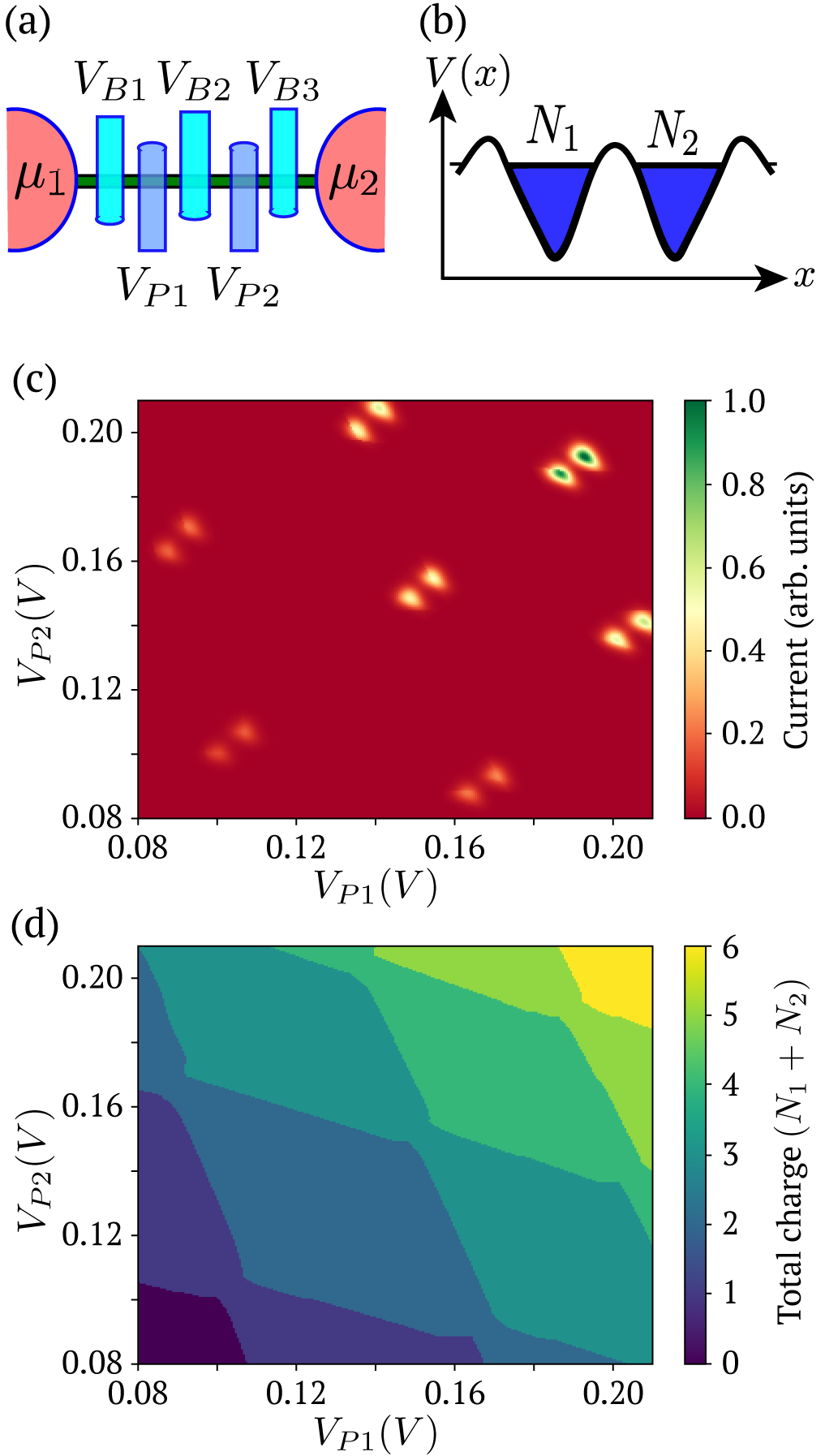}
\caption{(a) A 5-gate device used to model a double dot. (b) Potential profile $V(x)$ with charges $N_1$ and $N_2$ on the two dots. (c) Simulated current flow at triple points and (d) honey-comb charge stability diagram in the space of plunger gate voltages $(V_{P1},V_{P2})$ from the Thomas-Fermi model described in appendix~\ref{appendix:tf}}
\label{fig:5-gate-device}
\end{figure}

As a check on the qualitative performance of our model, we consider 3-gate and 5-gate configurations, as shown in Fig.~\ref{fig:3-gates-model}a and~\ref{fig:5-gate-device}a, respectively. We consider a single island (two islands) defined using three (five) electrostatic gates, 
$V_{Bi}$, with $i=1,2$ ($i=1,2,3$) and $V_{Pj}$ with $j=1$ ($j=1,2$). By changing the depth of the wells, electrons can tunnel in or out of the islands. At a given value of the gate voltage, a fixed integer number of electrons are assumed to exist on each island. Current flows through the device when two charge states have the same energy predicted by the capacitance model. In such a state, electrons tunnel through one of the contacts into the island (or tunnel between islands for a 5-gate device) and tunnel out of the island through another contact. The direction of the electron flow is set by the sign of the bias applied across the leads.

In the simulation for the 3-gate device (Fig.~\ref{fig:3-gates-model}(a) and~\ref{fig:3-gates-model}(b), a single dot is present along the nanowire. The contact chemical potentials are fixed to $\mu_1 = \mu_2 = 100.0\,$\si{\milli \electronvolt} with respect to the conduction band minimum. An infinitesimal bias of $10\,$\si{\micro \electronvolt} is present across the leads. The barrier gates are assumed to be kept at a fixed voltage with $V_{B1} = V_{B2} = -200\,$\si{\milli \volt}. The third gate, $V_P$, is swept from $0\,$\si{\milli \volt} to $350\,$\si{\milli \volt}.

As can be seen in the current trace in Fig.~\ref{fig:3-gates-model}(c), Coulomb blockade is reproduced. Our model also allows us to predict the most probable charge configuration from the Markov chain analysis. We see that the charge configuration jumps to a different state exactly at the position of the current peaks. The steady increase in the height of current peaks is a result of lowering of the tunnel barriers on increasing $V_P$. The decrease in spacing between adjacent current peaks with increasing values of $V_P$ is due to a slow increase in the capacitance of the dot with increasing electron number.

For the 5-gates device, (Fig.~\ref{fig:5-gate-device}(a)), the barrier voltages are set to $V_{B1} = V_{B2} = V_{B3} = -200\,$\si{\milli \volt}. These values were chosen so that the device operates in a double dot configuration (Fig.~\ref{fig:5-gate-device}(b)). We calculate the current as function of the two plunger gate voltages, $V_{P1}$ and $V_{P2}$. We reproduce the expected features for such a system~\cite{Wiel2003}, current flow only at triple points and honeycomb-shaped fixed-charge contours.(Fig.~\ref{fig:5-gate-device}(c) and~\ref{fig:5-gate-device}(d)).

We note that while more sophisticated models should be used in future studies, we find our approach to be sufficient for showing how ML can help with the challenges outlined in Sec.~\ref{sec:motivation}.

\section{Learning Coulomb blockade}\label{sec:learning-Coulomb-blockade}
We start our analysis from investigation of whether a machine can learn to identify the charge on a single quantum dot, given the current as a function of $V_P$ (Fig.~\ref{fig:3-gates-model}(c)). Formally, we define the broader problem of \emph{Learning Coulomb Blockade} as:

\begin{theorem}
\textbf{$\mathcal{P}1$: Charge Identification}\\
Given $I(\textbf{V})$, find a map $\mathcal{M}$ such that
\[ \mathcal{M} : I(\textbf{V}) \rightarrow \textbf{CS}(\textbf{V}), \]
where $I$ is the current at infinitesimal bias, \textbf{V} denotes the vector of voltages applied to the gates and \textbf{CS} is a vector of number of electrons on each island.
\end{theorem}

\begin{figure}[t]
\centering
\includegraphics[width=\linewidth]{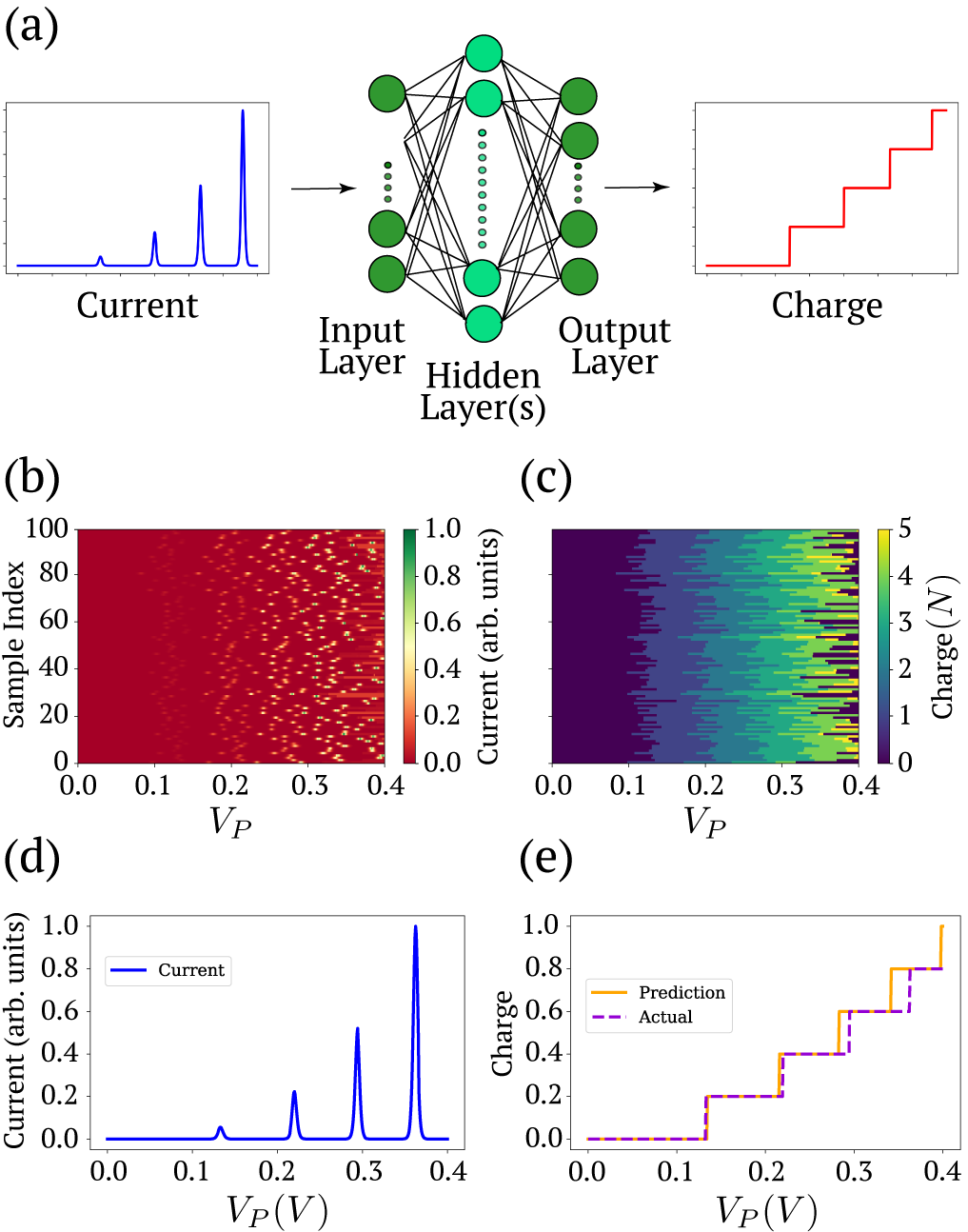}
\caption{(a) Overview of the ML problem of going from current to charge state for a single dot (b) Current vs $V_P$ data for 100 different dots. Each row represents a separate device with distinct gate positions and physical parameters sampled from a Gaussian distribution around a mean set of parameters (appendix~\ref{appendix:dataset}). (c) Corresponding charge vs $V_P$ data for current data from (b). (d) A sample current vs $V_P$ curve given as input to the trained DNN. (e) The output from the DNN showing the predicted and actual charge states for sample in (d).}
\label{fig:cb-learning}
\end{figure}

In the case of a single dot, only one gate voltage, $V_p$, is varied and the charge state is simply the number of electrons on the dot. Hence, $\textbf{V}$ and $\textbf{CS}$ are scalars.
It is easy to see that this just amounts to learning to integrate the current characteristics and scaling to the appropriate charge number (Fig.~\ref{fig:cb-learning}(a)).

We generated a training data set for 1000 distinct realizations of the dots. Each sample point is a current and charge state vs. $V_P$ characteristic. Across the samples, parameters such as the gate positions, widths and heights are sampled from a Gaussian distribution with mean values in the parameter set (standard deviation for the Gaussian was set to 0.05 times the mean value) (see 
Appendix~\ref{appendix:dataset} for details). Fig.~\ref{fig:cb-learning}(b) and~\ref{fig:cb-learning}(c) show sample current and charge data, respectively, of 100 such dots. The rationale behind generating a large dataset for the dots is twofold: having a variation in the dot parameters models the variations in different dots that are used in experiments and it presents a way to generate a generic training dataset for learning.

The machine learning problem is intended to map the current, given in Fig.~\ref{fig:cb-learning}(b), to the charge state, shown in Fig.\ref{fig:cb-learning}(c). One can think of this as a regression problem from the vector of current values to the vector of charge values.

We used a deep neural network with three hidden layers~\cite{Nielsen15} and achieved $91\%$ accuracy for the charge state values (see Appendix~\ref{appendix:tensorflow} for a description of the computing environment). Here, the accuracy for a single current-gate voltage curve (see Fig.~\ref{fig:cb-learning}(d) and Fig.~\ref{fig:cb-learning}(e)) is calculated from the predicted charge state from the neural network and the charge state from the Thomas-Fermi model over the gate-voltage range. This accuracy is them averaged over all the samples to produce an accuracy for the test set. The size of input and output layers correspond to the number of points in the $\textbf{I(V)}$ and $\textbf{CS(V)}$. We used a 512 point input and 512 point output layer. The result from the output layer was rounded to the nearest integer to get the charge state. The hidden layers comprised 1024, 256 and 12 neurons, respectively. The outcome of the training is a set of biases and weights corresponding to each neuron that allow the calculation of the final output.

Interestingly, we observed that a successive decrease in the number of neurons across the hidden layers
was critical to achieving a respectable accuracy. This suggests a redundancy of information encoded in the current characteristics that 
the network must learn to ignore when estimating the charge states.

We can visualize the learning by means of a validation set at the end of a fixed number of training epochs. We observed that in the initial training stages, the network learning the charge boundaries in the plunger voltage space of an \emph{average dot} and then slowly starts to learn to identify charge states of individual dot samples. 

We note that the problem identified above suffers from the charge-offset problem in the real world since the initial number of electrons on the dot might be unidentified. Hence, the network trained as a solution to Problem $\mathcal{P}$1 has limited applicability in experimental settings but nevertheless exemplifies that machine learning can, in principle, be applied to charge identification.

The charge number identification on the single dot also offers a trivial solution to identifying the state of the single dot. If the charge on the dot is non-zero, we can then conclude that a single dot exists whereas a zero charge implies a no dot device. The identification of state of the device with multiple islands from the current presents additional possibilities which we describe in the next section.

\section{Learning state}\label{sec:learning-state}

\begin{figure}[t]
\centering
\includegraphics[width=\linewidth]{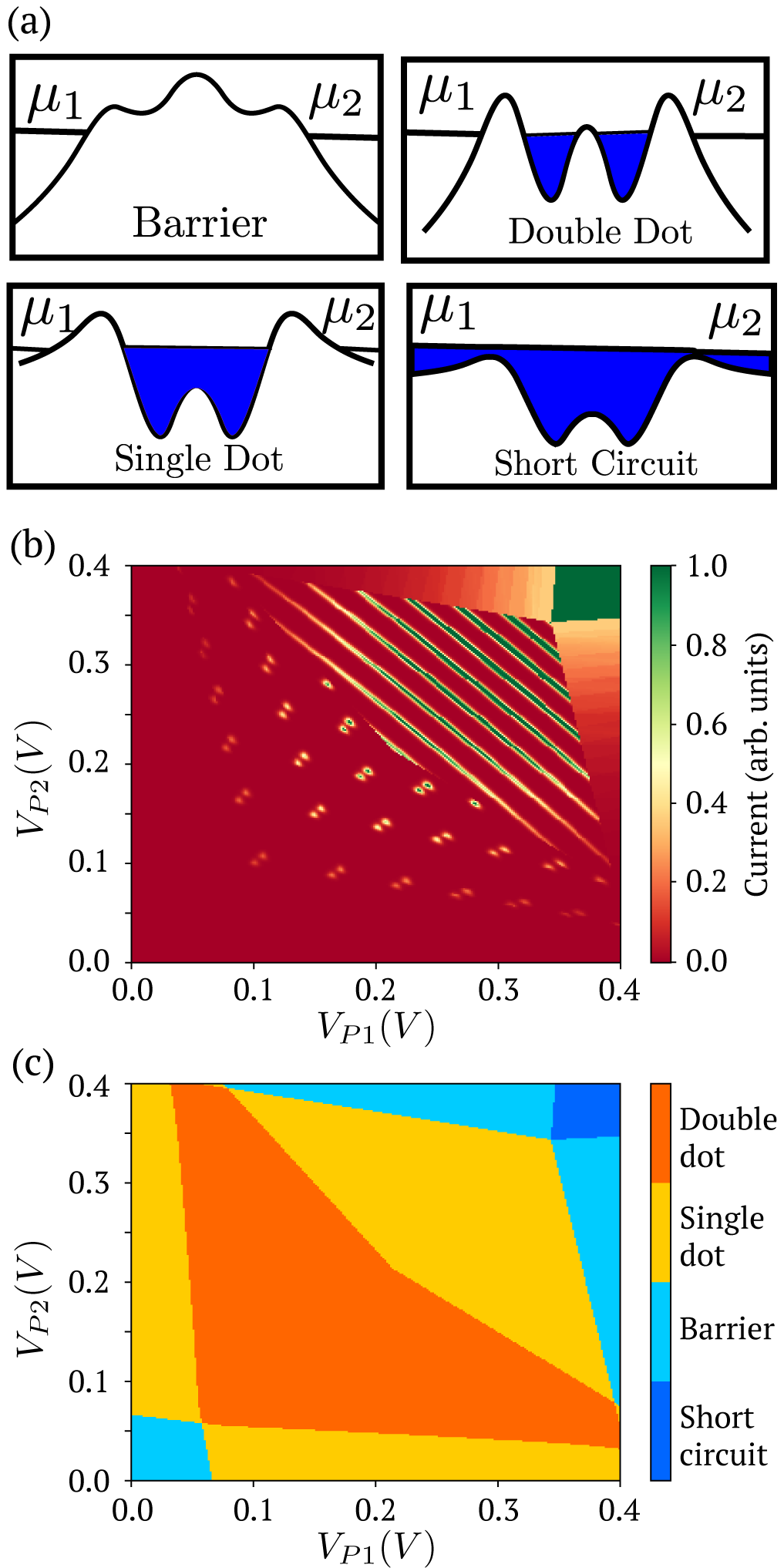}
\caption{(a) Possible states in the 5-gate device depending on the choices of the plunger gate voltages (b) Current vs $(V_{P1},V_{P2})$ exhibiting varied features in the current depending on the underlying state of the nanowire. (c) State vs  $(V_{P1},V_{P2})$}
\label{fig:dd-states}
\end{figure}

The \emph{state} is the number of distinct dots or islands that exist in the nanowire. We now consider a 5 gate device which can exist in 4 possible dot configurations: Quantum Point Contact or a Barrier, single dot (SD), double dot (DD) and a short circuit (SC) (see Fig.~\ref{fig:5-gate-device}(a) for the device model and Fig.~\ref{fig:dd-states}(a) for the possible dot configurations). Different states are reached by changing the voltages $V_{P1}$ and $V_{P2}$. The voltages $V_{B1}$, $V_{B2}$ and $V_{B3}$ are all fixed to $-200\,$\si{\milli \volt}.

To quantify the definition of a dot configuration, we define a probability vector $\bm{p}$ at each point in the $\bm{V}$ space. The elements of $\bm{p}$ correspond to the probability of being in each of the configurations as described above, i.e., $\bm{p}=(\text{SC}, \text{Barrier}, \text{SD}, \text{DD})$. For example, for a state of a single dot, the probability vector $\bm{p}=(0,0,1.0,0).$ 
For a region in $\bm{V}$ space, $\bm{p}$ is defined as the average of the probability vectors for the points in the region.

We are interested in determining the dot configuration (i.e., distinguishing between SC, Barrier, SD and DD) for a given set of barrier and plunger gate voltages. Formally, we define the problem as follows: 

\begin{theorem}
\textbf{$\mathcal{P}$2: State Identification for full region}\\
Given $\textbf{I(V)}$, 
find the probability vector $\bm{p}$ at each point in the given voltage space.
\end{theorem}

We generated a training set of 1000 gate configurations. Each sample point is the full two-dimensional map ($100\times100$ points) from the space of plunger gate voltages $(V_{P1},V_{P2})$ to current (see Fig.~\ref{fig:dd-states}(b) for an example of such map). A state map corresponding the the current map presented in Fig.~\ref{fig:dd-states}(b) is shown in Fig.~\ref{fig:dd-states}(c). The states are calculated via the electron density predicted from Thomas-Fermi model. The number of distinct charge islands in the electron density separated by regions of zero electron density corresponding to the barriers is used to infer the state of the nanowire (see Appendix~\ref{appendix:tf}). Note that there is more than one way for some of the configurations to exist. For instance, lower voltages on the barrier B2 with respect to the barriers B1 and B3 or higher voltages on B1 and B2 as compared to B3; all lead to a single dot configuration (see Fig.~\ref{fig:5-gate-device}(a)). Analogously to the single dot case, gate and physical parameters are sampled from a Gaussian distribution with mean values in parameter set (see Appendix~\ref{appendix:dataset}).

We note that Problem $\mathcal{P}$2 is a regression problem from the $\bm{I(V)}$ space to the space of probability vectors. The aim is to go from Fig.~\ref{fig:dd-states}(b) to Fig.~\ref{fig:dd-states}(c). We used a similar neural network with three hidden layers that we employed for the single dot problem. The input and output layers are now of the size equal to number of points in the $\bm{I(V)}$ and $\bm{CS(V)}$ relationships, i.e., $100 \times 100$ points. It was possible to achieve $91\,\%$ on state values i.e., it was possible to reproduce the state map in Fig.~\ref{fig:dd-states}(c) across different devices with the state label agreeing to $91\,\%$ with the actual values. 

As far as tuning the device is considered, it is not very useful to know to probability vector at each point in the voltage space. Hence, we move to defining a probability vector for a sub-region as opposed to a single point in the voltage space.

\section{Auto-tuning}\label{sec:auto-tuning}
We define the process of finding a range of gate voltage values in which the device is in a specific configuration as \emph{auto-tuning}. The ability to characterize the state at any point in the voltage subspace provides a promising starting point for the automated tuning of the device to a particular state. In particular, having an automated protocol for achieving stable desired electron state would allow for efficient control and manipulation of the few electron configurations. In practice, auto-tuning compromises of two steps: (i) identifying the current state of the device and (ii) optimizing the voltage configuration to achieve a desired state. The steps are then repeated until the expected state is reached. For a device with $m$ gates, this leads to a problem of finding a $m$ dimensional cuboid in the space of the $m$ gate voltages. From a machine learning perspective, the recognition and tuning of the state can be expressed as the following two problems:

\begin{theorem}
\textbf{$\mathcal{P}$3a: State Identification for sub-region}
Given $\textbf{I(V)}$, find the average probability vector of the region.
\end{theorem}

\begin{theorem}
\textbf{$\mathcal{P}3$b: Auto-tuning}:
Given the $I(\textbf{V})$ characteristics, an initial subregion in $\textbf{V}$ space and a desired dot configuration, find (tune to) a subregion with the desired dot configuration. 
\end{theorem}

\begin{figure}[t]
\centering
\includegraphics[width=\linewidth]{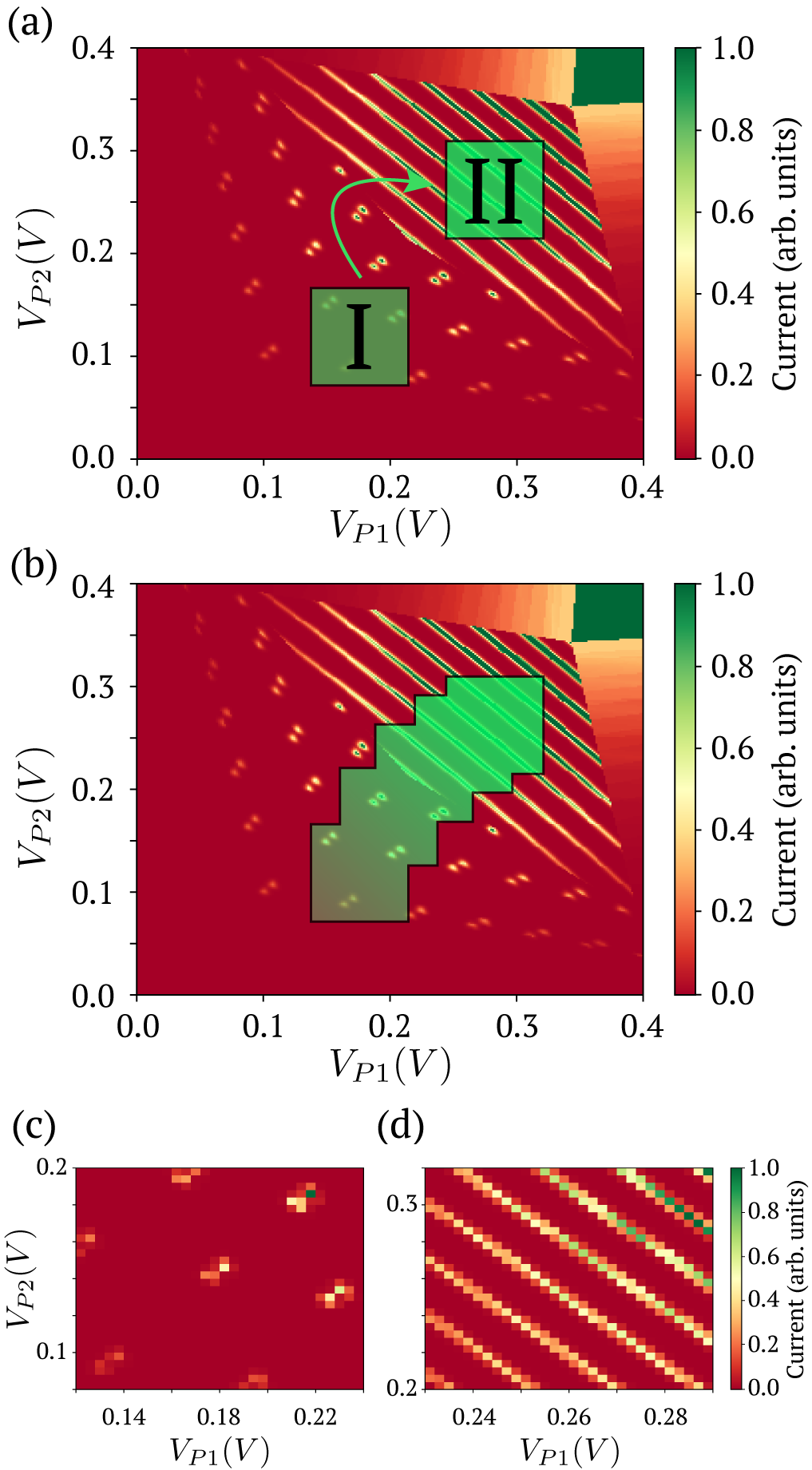}
\caption{(a) Idea behind auto tuning with I and II being the starting and ending sub regions respectively.  (b) Sub-regions encountered by the optimizer when auto-tuning to the single dot region, i.e., the destination probability vector $p_0$ being set to $(0,0,1,0)$. (c) Starting sub region (d) End sub region}
\label{fig:autotuning}
\end{figure}

The idea behind auto-tuning in a two-dimensional space is presented in Fig.~\ref{fig:autotuning}. For the case of 5-gate double dot device, defined in Sec.~\ref{sec:learning-state}, we consider the restricted problem with two gates $V_{P1}$ and $V_{P2}$ being controlled and the barrier gates remained fixed (see Fig.~\ref{fig:5-gate-device}(a)). We start out in a double dot region and the desired dot configuration is set to be a single dot region.

\subsection{State learning}\label{subsec:dev_learn}

As mentioned earlier, the first step in the auto-tuning process is the recognition of the existing configuration of the device. In a typical experiment one has access only to a limited voltage regime, decided upon by the experimentalist. Such a region can be thought of as a sub-image of the two-dimensional gate voltage map mentioned in Sec.~\ref{sec:learning-state}. The identification of the state of the device is an image classification task, with categories representing the different states of the nanowire (i.e., SC, Barrier, SD and DD). 

\begin{figure}[t]
\centering
\includegraphics[width=\linewidth]{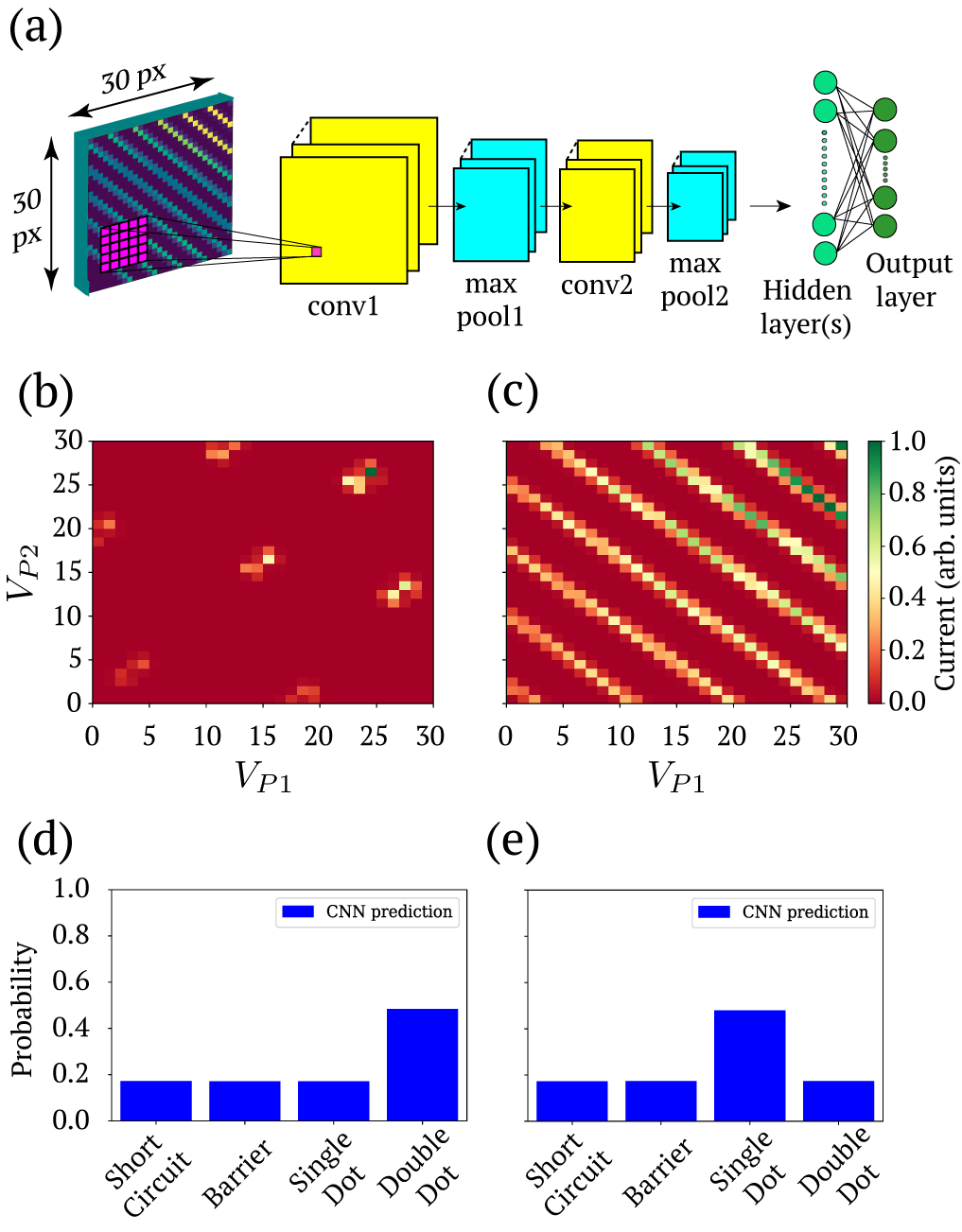}
\caption{(a) Design of the CNN for sub-region identification. $30 \times 30$ pixel images are used as input to the CNN (b) Current data and (d) probability vector predicted when sub-region is of double dot type. (c) Current data and (e) probability vector predicted when sub-region of single dot type. The axes ticks denote the pixel index.}
\label{fig:cnn-training}
\end{figure}

Such problems have been successfully solved by convolutional neural networks (CNNs). CNNs have one or more sets of convolutional and pooling layers, that precede the series of hidden layers (see Fig.~\ref{fig:cnn-training}(a)). A convolutional layer consists of a number of fixed size kernels which are convolved with the input. The number of kernels in a layer is referred to as the number of features in that layer. The weights in the kernel are determined by the training on the dataset. In order to reduce dimensionality of the input for faster operation and to effectively learn larger scale features in the input, a convolutional layer is generally followed by a pooling layer. A pooling layer takes in a sub-region in the input and replaces it by an effective element in that region. A common pooling strategy is to let the effective element be the maximum element in the sub-region which leads to the notion of a max-pooling layer.

The training set for the voltage subspace learning was generated based on the set of 1000 full two-dimensional maps of $I$ vs $(V_{P1},V_{P2})$ from Sec.~\ref{sec:learning-state}. 50\,000 sub-maps of a fixed size ($30\times30$ pixel) were generated. 90\,\% of the 50\,000 samples were used as the training set and the rest were used to evaluate the performance of the network. The network achieved 96\,\% accuracy in prediction of the state. Two examples of the sub-maps and corresponding probability vectors from the evaluation stage are presented in Fig.~\ref{fig:cnn-training}(b), (c) and Fig.~\ref{fig:cnn-training}(d), (e) respectively.

For the training, we used two convolutional layers with kernels of size $[5,5]$. The layers both had 16 features. Each convolutional layer was followed by a max-pooling layer, wherein the pool size was set to $[2,2]$. The two hidden layers consisted of 1024 and 256 neurons. Rectified linear units (ReLU) with a dropout rate of 0.5 were used as neurons. Dropout regularization  was introduced to avoid over-fitting~\cite{Hinton12}. Finally, an Adam optimizer was used to speed up the training process~\cite{Diederik14}.

We found that the introduction of the convolutional layers was crucial in achieving better results in terms of both accuracy and efficiency. Here, accuracy is defined with the prediction of the state with the highest probability and efficiency is defined in terms of training time. We note that the state is predicted from the highest probability, though it might be possible that this highest probability is less than 0.5 (see Fig.~\ref{fig:cnn-training}(d) and Fig.~\ref{fig:cnn-training}(e)). Introducing more hidden layers did not affect the accuracy as much as introducing convolutional layers; this indicates that classification over the features seems to be a simpler task than producing an effective representation for the features. 

\subsection{Tuning the device}\label{sec:sub_tuning}

Having the state of the device identified for a sub-region, the procedure of auto-tuning corresponds to a simple optimization problem. Let $\bm{p}$ be the probability vector of a given sub-region and $\bm{p}_0$ be the desired probability vector. Define $\delta(\bm{p},\bm{p}_0) = |\bm{p} - \bm{p}_0|$, where $|\cdot|$ denotes the vector norm. The problem of auto-tuning is then equivalent to minimization of $\delta(\bm{p},\bm{p}_0)$ over the space of gate voltages $\textbf{V}$.

We used COBYLA from the Python package SciPy~\cite{powell_1998} as a numerical optimizer. The probability vector $\bm{p}$ was calculated using the neural network described in Sec.~\ref{subsec:dev_learn}. The starting region was set initially in a double dot region, as can be seen in Fig.~\ref{fig:autotuning}(c). Around 15-30 evaluations of the probability vector using the CNN were required to ultimately find the required sub-region (Fig.~\ref{fig:autotuning}(d), II in Fig.~\ref{fig:autotuning}(a)) depending on the position of the initial subregion. The starting region was varied over the space of $(V_{P1},V_{P2})$ and in each case it was possible to auto-tune to the required sub region.

\section{Working with Experimental Data}\label{sec:experimental}

We ran the CNN with the set of weights and biases established during the training on the simulated dataset described in Sec.~\ref{subsec:dev_learn} on an experimental dataset for a 3-gate device from our group~\footnote{Location of experimental data: \url{Elwood:\\internal\\SET\_data\\wet\_DR\\Measurement Diagnostics\\12\_4 N R\\data}; files are May1\_1 to \_16.dat and Apr24\_1 to \_49.dat.}; 
the device is as described in~\cite{Koppinen2013}, and the measurements are very similar to those presented in \cite{Fujiwara2004}. The device used in the experiment had two barrier gates (B1 and B2) and one plunger gate (P). In this device, the barrier gates were also used as generic plunger gates. By choosing appropriate voltage values for the plunger (P) gate, the device could be operated as a single dot or a double dot device. The measured data consisted of 2D differential conductance maps in the space of barrier gate voltages ($V_{B1}$ and $V_{B2}$) for varied but fixed values of the plunger voltage ($V_P$). Since the qualitative features are similar for a current map and a differential conductance map, we could feed in the differential conductance output into the CNN. 

\subsection{Identification of state in experimental Data}

\begin{figure}[t]
\centering
\includegraphics[width=\linewidth]{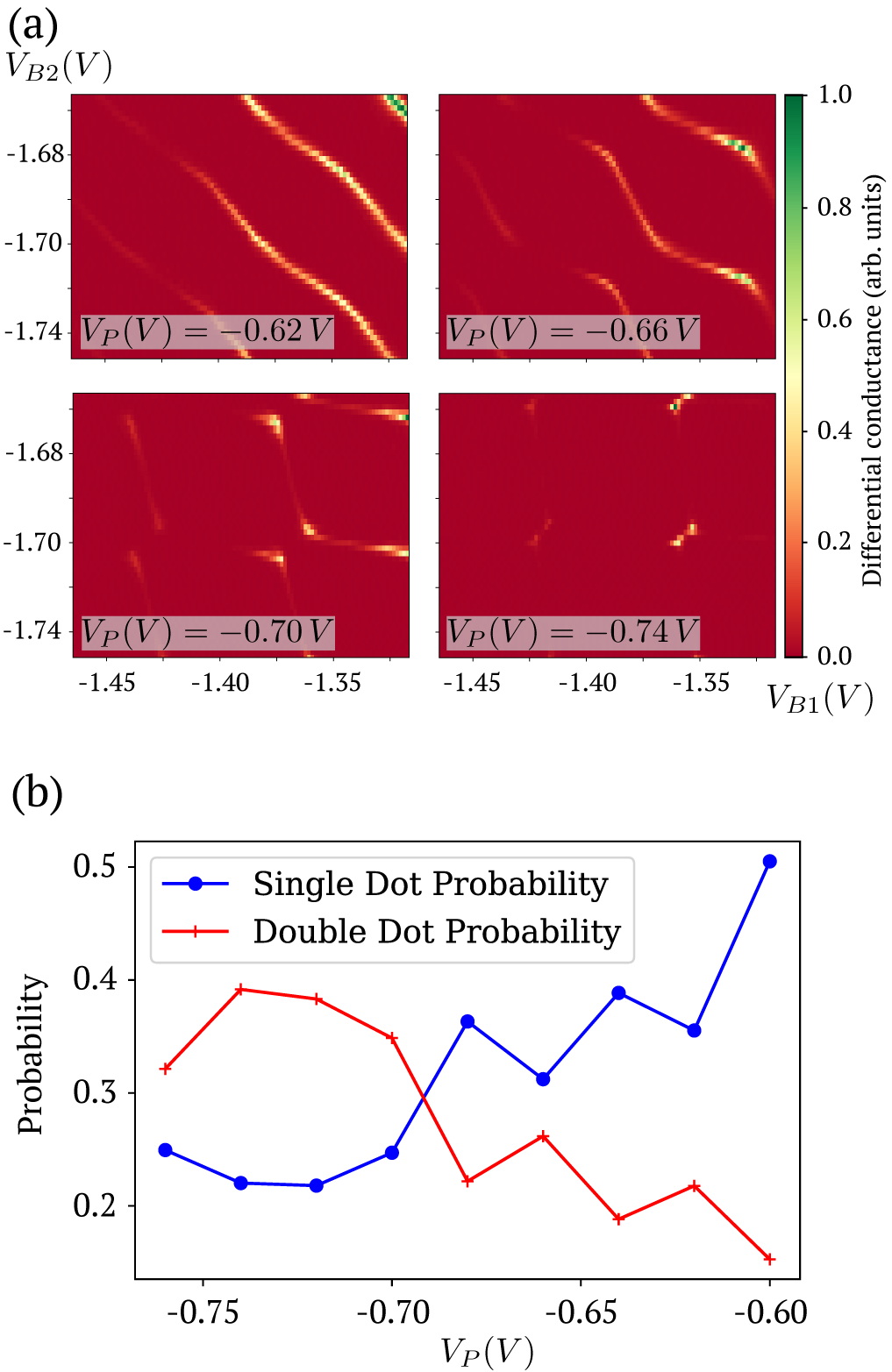}
\caption{(a) Experimental data at different values of the plunger gate exhibiting a transition from a single dot state to a double dot state. (b) The predicted state probability by the CNN as a function of $V_p$. A clear transition is seen from single dot state to double dot state as is intuitively seen in the experimental data. The probabilities predicted from the CNN for the other states (barrier and short circuit) are smaller in comparison to the single or double dot probabilities and hence only the highest two probabilities corresponding to these states are shown for clarity.}
\label{fig:exp-data-probs}
\end{figure}

For state identification, we considered small regions in the space of barrier voltage for a fixed plunger so that in each of the maps the device was in only one of the states, single or double dot. The maps were then taken at different values of the plunger voltage, ranging from $-0.76\,$\si{\volt} to $-0.60\,$\si{\volt}. The barrier gates are varied 
from $-1.44\,$\si{\volt} to $-1.34\,$\si{\volt}. Fig.~\ref{fig:exp-data-probs}(a) shows the 2D maps for different values of the plunger voltage. A gradual transition from a single dot device to a double dot device is seen.

Since our model produces the current value only qualitatively, the experimental data had to be re-scaled (by a constant number) prior to feeding into the CNN to match the simulated data. The CNN characterizes the state present in the device through a probability vector. Results for different values of the plunger voltage are shown in Fig.~\ref{fig:exp-data-probs}(b). As can be seen, our CNN can effectively distinguish a single dot and a double dot state from the current maps. 

\subsection{Auto-tuning of the device to a double dot state}
\begin{figure}[t]
\centering
\includegraphics[width=\linewidth]{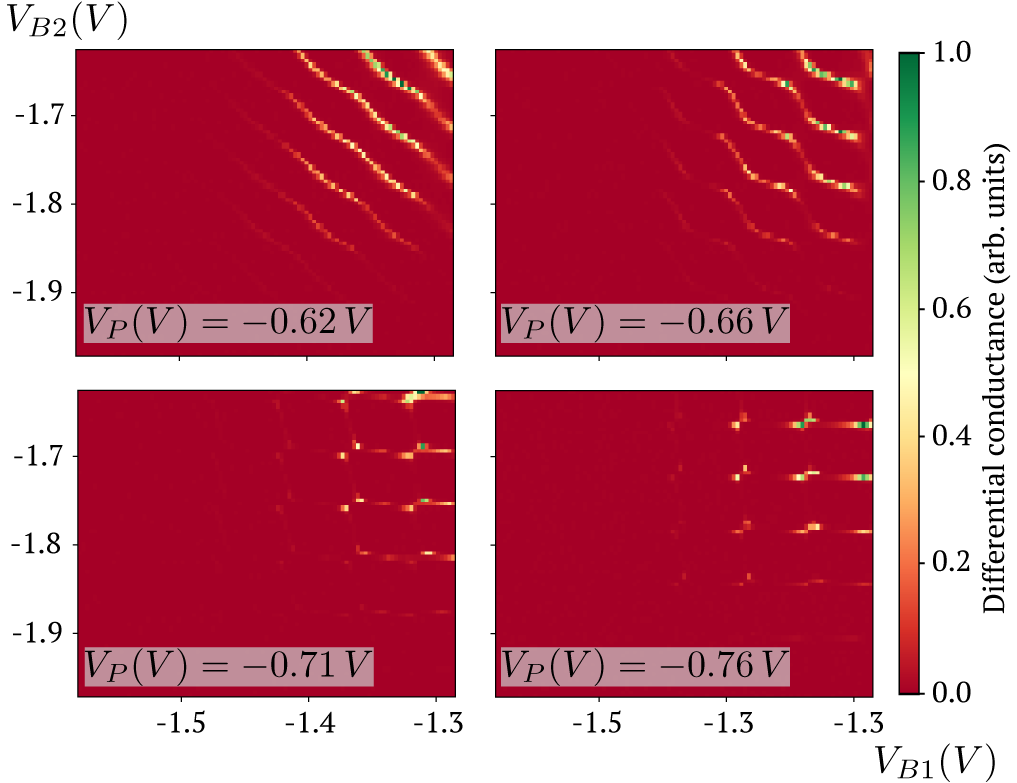}
\caption{Experimental data from the same device with a wider variation in the barrier gate voltages. Differential conductance is measured as a function of the barrier gate voltages ($V_{B1}$ and $V_{B2}$) and is plotted for four different values of the plunger voltage. The device shows a gradual transition from a a single dot current characteristics to a double dot current characteristics for more negative values of the plunger voltage.}
\label{fig:full_exp}
\end{figure}

Since the device state could be predicted with reasonable accuracy, we considered tuning gate voltages from one state to another based on the experimental data. For this part, a dataset with a larger variation in barrier voltages was used. Figure~\ref{fig:full_exp} shows 2D maps of differential conductance vs the barrier gate voltages ($V_{B1}$ and $V_{B2}$) for four different values of the plunger voltage. 2D sub-regions of these maps were used as input to the CNN in the tuning procedure.

\begin{figure}[t]
\begin{minipage}{\linewidth}
\centering
\includegraphics[width=\linewidth]{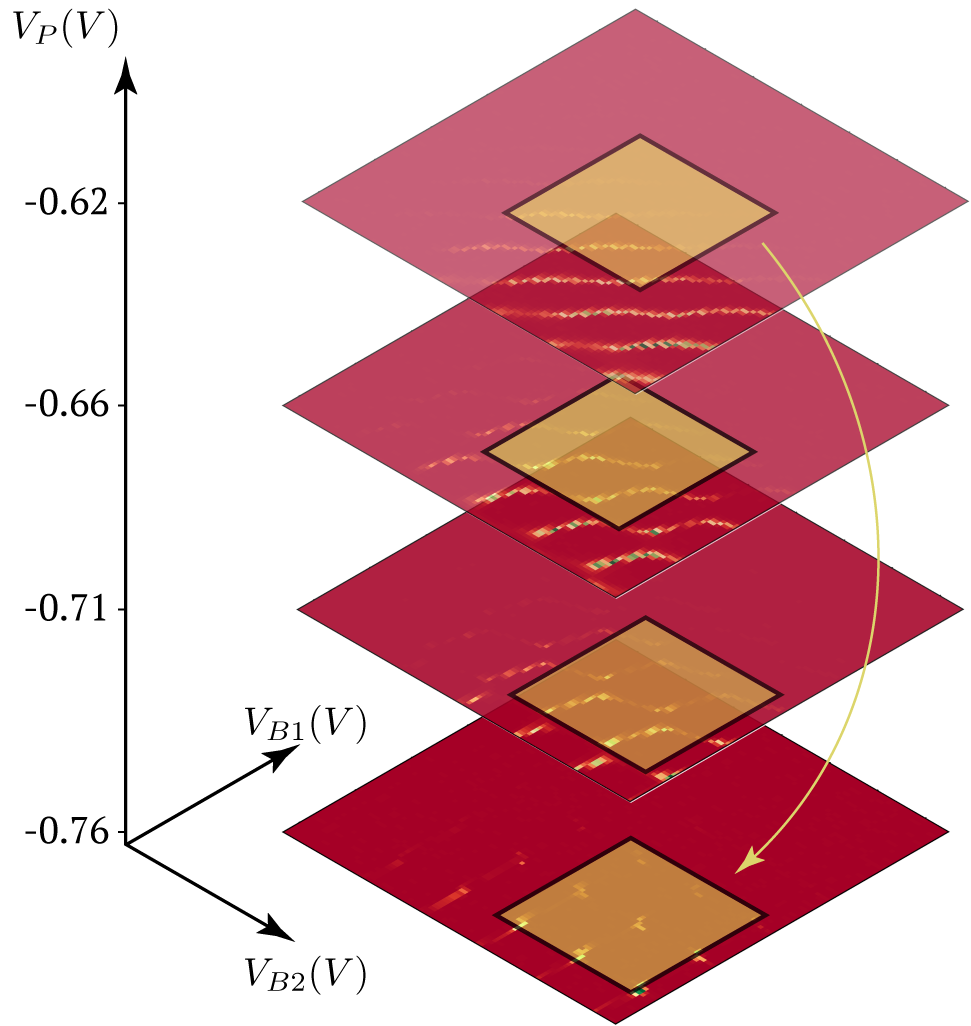}
\end{minipage}
\caption{The idea behind auto-tuning in the three dimensional space of two barrier and plunger gate voltages. The successive squares represent the sub-regions encountered in the tuning process which are fed as input to the CNN. The arrow represents the direction of movement in going from an initial region to a final region. See Fig.~\ref{fig:full_exp} for the $V_{B1}$ and $V_{B2}$ range.}\label{fig:tuner}
\end{figure}

\begin{figure}
\centering
\includegraphics[width=\linewidth]{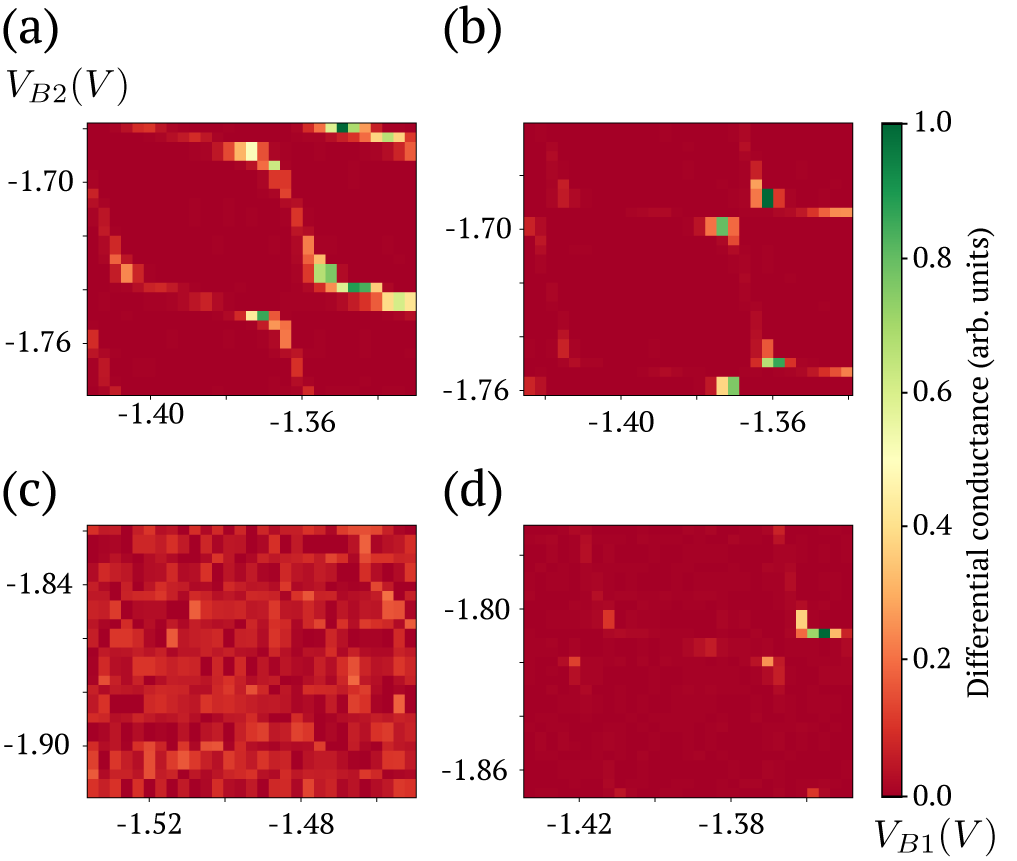}
\caption{(a) The initial region ($V_P = -0.64\,$\si{\volt}) set in a single dot region. (b) The optimizer coupled with the CNN tunes the device to a double dot state ($V_P = -0.73\,$\si{\volt}) i.e. the destination probability vector $p_0$ being set to $(0,0,0,1)$. (c) The initial region set to a region ($V_P = -0.66\,$\si{\volt}) with no current through the device.(d) The final state is again a double dot ($V_P = -0.72\,$\si{\volt}) as required.}\label{fig:exp_tuning}
\end{figure}

We considered the auto-tuning of all three gate voltages (two barriers and the plunger).  The final tuned state was set to a double dot region. See Fig.~\ref{fig:tuner} for a visualization of the auto-tuning process. Two kinds of initial regions were considered: a single dot region (Fig.~\ref{fig:exp_tuning}(a)) and a region with no current (Fig.~\ref{fig:exp_tuning}(c)). In both cases, it was possible to find a set of barrier and plunger gate voltages that map to a double dot state (Fig.~\ref{fig:exp_tuning}(b) and Fig.~\ref{fig:exp_tuning}(d)).  Effectively, the CNN predicted the probability vector describing the device state (sec.~\ref{sec:learning-state}) from maps at different plunger voltages and the optimizer tuned the probability vector to a required form (in this case, a double dot).

We used the same optimizer as described in Sec.~\ref{sec:sub_tuning}. The tuning process was  completed within 40 to 50 iterations, depending on the initial region. Hence, the CNN coupled with an optimizer can be used with data from actual experiments for auto-tuning the device state.

\section{Discussion}\label{sec:discussion}
\subsection{Neural Networks in an Experimental Setting}

We describe how a generalized auto-tuner neural network can be implemented in an experiment to automatically adjust the parameters of the device to an expected state. Consider a quantum dot device with a set of gate voltages $\bm{V}$. We showed that a neural network can be trained to predict a probability vector $\bm{p}$ describing the state of an arbitrary sub-region in the $\bm{V}$ space. This predicted vector $\bm{p}$, together with a destination probability vector, can be then fed to an optimizer controlling the space parameters in order to obtained the desired single or double dot state. 

In particular, let's assume that $\bm{p}_0$ is the probability vector of the desired state. Starting in a random region of the voltage space, the trained CNN can predict a probability vector $\bm{p}$ for this region. A fitness function $\delta$ is then used to compare the predicted probability vector $\bm{p}$ and the destination vector $p_0$. By minimizing $\delta$, the auto-tuning of the device takes place. An optimizer determines an optimal set of parameters that leads to a new sub-region. The process is then repeated until the fitness has been minimized to a particular value.

Since, the entire voltage space in $\bm{V}$ does not have to be explored, this implies a saving in terms of experimental time. Also the process does not use human intervention at any step in the tuning of the dot signifying the use of \emph{auto} in our definition of the auto-tuning problem.

\subsection{Further Problems}
We have presented novel techniques towards tuning of quantum dot devices. Given that building scalable quantum computing devices is now on the horizon, we hope that such methods will present themselves as natural subroutines for construction of real devices and will do away the need to rely on heuristics. Hence, we outline further problems that are more realistic and useful in an experimental setting and can be potentially tackled with machine learning.
\begin{theorem}
$\mathcal{P}4$: \textbf{Inductive Learning}

\end{theorem}
Moving to learning and auto-tuning of multiple dots will present new challenges as a result of the higher dimensional space of gate voltages. This curse of dimensionality might detrimentally affect the design of auto-tuning algorithms. Pattern recognition in dimensions greater than 2 has not been studied extensively. Instead we propose a different solution that can be generalized based on an inductive strategy. We refer to it as \emph{inductive learning}.

In \emph{Inductive Learning}, we make use of the fact that gates which are spatially far apart are likely to be loosely coupled to each other. Hence, a strategy emerges in which we use the auto-tuning algorithm to tune the first two barrier gates. A second type of neural network will be used to tune the plunger gate. This will be repeated until all the single dots formed by 2 barriers and a plunger are tuned to the required configurations.

\begin{theorem}
$\mathcal{P}5$: \textbf{Charge Tuning}
\end{theorem}

The capacitance matrix is an effective model of the device and it determines the quantitative size of features in the current output. For instance, in the case of a single dot, the capacitance matrix can be directly related to the charging energy of the device. For the double dot, the capacitance matrix elements determine the size of the honeycomb hexagons. Hence, establishing a learning algorithm for the capacitance matrix is the next logical step. A capacitance matrix along with the voltage values of the gate can be used to estimate the charge on the device. Estimation of the charge can then be coupled with an optimizer to tune the device to required charge values exactly like tuning the state as described in this paper.  We refer to this process of learning the capacitance matrix and tuning the charge on the device as \emph{Charge Tuning}.

We remark here that these further problems and any other that might arise may require different types of machine learning algorithms beyond just deep and convolutional neural networks described in this paper.

\section{Conclusion}\label{sec:conclusion}
We have described a bare-bones physical model to calculate the capacitance matrix for a linear array of gate defined quantum dots. We used a Markov chain model amongst the charge to simulate transport characteristics under infinitesimal bias. Our model can qualitatively reproduce the current vs gate voltage characteristics observed in experiments. 

This model was used to train deep neural networks to learn the charge and state of single quantum dots from their current characteristics. We used a convolutional neural network to identify state of a double quantum dot device from two-dimensional current maps in the space of gate voltages. We defined the auto-tuning problem for quantum dot devices and described strategies for tuning single and double dot devices. The trained networks were tested on experimental data and successfully distinguished the single and double dot device states. We also demonstrated auto-tuning in a three-dimensional space of barrier \& gate voltages on an experimental dataset. 

Finally, we described how an auto-tuner network might be incorporated in an experiment and outlined further problems in tuning of quantum dot devices. Moreover, our work presents an example of machine learning techniques, specifically convolutional neural networks, fruitfully applied to experiments, thereby paving a path for similar approaches to a wide range of experiments in physics.

\begin{acknowledgements}
We thank Eric Shirley and Michael Gullans of NIST for helpful discussions. SSK acknowledges financial support from the S. N. Bose Fellowship. We acknowledge funding from the NSF Physics Frontier Center at the JQI and the Army Research Laboratory funded CDQI. Any mention of commercial products is for information only; it does not imply recommendation or endorsement by NIST.
\end{acknowledgements}

\appendix
\section{Thomas-Fermi Capacitance Model}
\label{appendix:tf}
\subsection{Calculation of electron density under Thomas-Fermi approximation}
We model the electron density as an inhomogeneous electron gas originally used in the statistical theory of Thomas and Fermi for atoms \cite{March}. In this theory, properties of a homogeneous electron gas are applied locally to the inhomogeneous electron gas. This assumption is referred to as the Thomas-Fermi (TF) approximation and is justified when the electron density or the potential acting on it do not change appreciably over a characteristic electron wavelength.

An externally created potential $V(\bm{x})$, e.g. from gates, is assumed to be given. Electron density $n(\bm{x})$ is treated as the dynamical variable to be found in the theory. A Fermi level $\mu_F$ is given for the electron gas. For the purposes of simulations presented in this paper, we assume a electron density $n(x)$ on a finite one dimensional grid. (Fig.~\ref{fig:nanowire_model}(b))

Consider a Fermi sea with Fermi energy $\mu_F$. Let the bottom of the conduction band be at energy $\epsilon_0$. In the absence of an external potential, the electron density in the conduction band can be calculated as,

\begin{equation}\label{eqn:elec-density}
    n = \int_{\epsilon_0}^{\infty} \frac{g(\epsilon)}{1 + e^{\beta(\epsilon - E_F)}} d\epsilon 
\end{equation}
where $g(\epsilon)$ is the density of states in the conduction band and $\beta$ is the inverse temperature.

Due to the presence of an external potential, the conduction band minimum shifts in energy. Moreover, the electron density produces an effective potential due to the Coulomb self-interaction. As a result, the band minimum is modified as,

\begin{equation}\label{eqn:mod-band-min}
    \epsilon_0'(x) = \epsilon_0 - e V(x) + \int K(x,x') n(x') dx'
\end{equation}
where $\epsilon_0'(x)$ is the new spatially varying band minimum, V(x) is the externally applied potential and $K(x,x') =\frac{K_0}{\sqrt{(x-x')^2 + \sigma^2}}$ gives the Coulomb energy between points $x$ and $x'$. $K_0$ sets the energy scale of the interaction. A softening parameter $\sigma$ has been added to the denominator and serves a twofold purpose: it models the effective one-dimensional interaction for a higher dimensional gas of electrons as would be present in the device and prevents a numerical singularity at $x=x'$. $\int K(x,x') n(x') dx'$ gives the effective Coulomb potential created as a result of the electron density $n(x)$. Since the effects of the electron density on the conduction band minimum are also included, equation~\ref{eqn:elec-density} with the modified band minimum, equation~\ref{eqn:mod-band-min}, provide a self-consistent calculation of the electron density $n(x)$.

In our calculations, we assume a two-dimensional electron gas (2DEG) to model the electron density of states. The density of states for a 2DEG, $g(\epsilon) = g_0 = \frac{m*}{\pi \hbar^2}$ is equal to a constant. Equation~\ref{eqn:elec-density} was solved in an iterative fashion. The starting solution was taken as $n(x) = 0$ which was plugged in~\ref{eqn:mod-band-min}. The modified band minimum was then used to calculate the $n(x)$ using~\ref{eqn:elec-density}. This iteration was repeated until the density $n(x)$ converged. The strength of the Coulomb interaction was increased in a linear fashion to its required strength for a fixed initial number of iterations to avoid pathologies associated with numerical convergence in the self-consistent calculation. 

The device is assumed to be connected to large reservoirs of electrons present as the contacts. The contacts are assumed to be kept at a fixed and equal chemical potential $\mu = \mu_F$. As an approximation, in the absence of gate potentials the conduction band minimum of the entire one dimensional device is assumed to be a constant function of x, being equal to chemical potential of the contacts. Intuitively, the points where $V(x) = \mu_F$ are the classical turning points for the electrons and differentiate regions of islands and barriers. The regions where $\mu > V(x)$ constitute islands of electrons and the rest where $\mu < V(x)$ (classically forbidden regions) as forming barrier regions between islands.

\subsection{Calculation of a capacitance model}
Consider a system of $m$ conductors. A capacitance can be defined between each conductor and every other conductor as well as a capacitance from each conductor to ground. The relation between charges on the islands and their electrostatic potential can then be conveniently expressed with a capacitance matrix $\mathbb{C}$ of size $m \times m$.
\begin{equation}
\bm{Q} = \mathbb{C} \bm{V}
\end{equation}
$\bm{Q}$ is the vector of charges on each conductor and $\bm{V}$ is a vector with the voltage on each conductor with respect to a ground potential. The conductors are coupled capacitively to fixed voltages which act as gates in the actual device. The electrostatic energy $E$ of the system of conductors can be expressed as:
\begin{equation}
E = \frac{1}{2} (\bm{Q - Z})^{T} \mathbb{C}^{-1} (\bm {Q - Z})
\end{equation}
where $\bm{Z}$ is the vector of induced charges due to the gates. 

The physics of transport in electrostatically coupled quantum dots with negligible inter-dot tunnel conductance can be described by an orthodox Coulomb blockade theory \cite{Wiel2003}. We work with a purely classical description of the electron density islands in our one dimensional system without the inclusion of discrete quantum states. We regard them as a system of conductors having a \emph{discrete} number of electrons and influencing the charges on each other via a capacitance matrix.

A capacitance model of the system is defined as tuple $(\mathbb{C},\bm{Z})$ where $\mathbb{C}$ is the capacitance matrix of the islands and $\bm{Z}$ is the vector of induced charges. We wish to establish a procedure to calculate a capacitance model for islands formed in our system. 

Electron density $n(x)$ is calculated using equation~\ref{eqn:elec-density}. Assume that the electron density is such that it is non-zero in certain regions (the islands) and zero between the islands (Fig.\ref{fig:nanowire_model}(b)). $\bm{Z}$ is then calculated by integrating the electron density over each island and is treated as the charge induced by the gate potentials in the capacitance model.
 
Let $\bm{Q}$ be the vector of charges on each island. Since the number of electrons on each island is assumed to be an integer, each element of $\bm{Q}$ is an integer times the electronic charge as opposed to elements of $\bm{Z}$ which can take arbitrary real values. The energy $E$ of a charge configuration is given as:
 \begin{align}\label{eqn:cap-model}
 E &= (\bm{Q}-\bm{Z})^{T}\left(\frac{1}{2 \mathbb{C}}\right)^{-1}(\bm{Q}-\bm{Z}) \\
   &= \sum_{i,j} E_{i,j} (\bm{Q}-\bm{Z})_i (\bm{Q}-\bm{Z})_j
 \end{align}
where $\left(\frac{1}{2\mathcal{C}}\right)^{-1}_{i,j} = {E_{i,j}}$.

The energy calculated using this capacitance model is a manifestation of the kinetic energy of the Fermi sea in each island and the Coulomb interaction between the islands. We can use this correspondence to calculate the inverse capacitance matrix elements $E_{i,j}$ using the Coulomb interaction potential $K(x,x')$ and the electron density $n(x)$.

\begin{align}
E_{i,j} = \frac{c_k \delta_{i,j} \int_{i}n(x)^2 dx + \int_{i}\int_{j}K(x,x') n(x)n(x') dx}{\left( \int_{i}n(x) dx\right)\left( \int_{j}n(x) dx\right) }
\end{align}
where $ \delta_{i,j}$ is the Kroneckor delta function and $c_k$ is the coefficient which sets the scale for kinetic energy of the Fermi sea in each island. The integration subscript $\int_i$ denotes that the integration is to be performed only over the extent of the $i^{\text{th}}$ island. The denominator has been added to normalize with the total number of electrons on each island.

Determination of the elements $E_{i,j}$ and $\bm{Z}$ amount to determination of the capacitance model for the islands.

\subsection{Calculation of equilibrium charge distribution}
Once the capacitance model has been calculated, we calculate energies of charge configurations closest to the induced charges values, $\bm{Z}$, while constraining the number of electrons on each island to be integers. The equilibrium charge configuration is set to the one with the lowest energy.

\section{Markov Chain Model for mesoscopic transport in the semi-classical regime}\label{appendix:markov}
In order to simulate transport characteristics and calculate a current given a potential profile $V(x)$, we introduce a Markov chain model. The actual physics included in this abstract model is calculated using the Thomas-Fermi approximation defined in appendix ~\ref{appendix:tf}. We assume that the contacts are kept under an infinitesimal bias so that the current flow can be modeled by an elastic tunneling Hamiltonian. The tunnel rates are estimated using the WKB approximation.

\subsection{Graph definitions and construction}
Consider a system with $k$ islands where each island is assumed to have $N_{di}$ electrons ($i = 0,1,2, ..., k)$. $\bm{N}_d = (N_{d1},...,N_{dk})$ is referred to as a charge configuration of the islands. Let $G= (V,E)$ be a \emph{directed} graph. Each node $v \in V$ is a charge configuration as defined above. An edge exists between two nodes if they are connected by an electron tunneling event, either across adjacent islands or through the leads. We introduce an order $p$ of the graph model which is defined such that $|N_{di} - \mathbb{Z}_i| \leq p \quad \forall i=1, .., k$. 

Each graph is constructed in a breadth-first fashion from a starting node. The charge configuration $\bm{Z}$ as defined in appendix~\ref{appendix:tf} is used as a starting node for the constructing the graph. All charge configurations which can be reached from this state by a single electron tunneling event are found and are added to the set of nodes. This procedure is recursively repeated for the new nodes added to the graph until no new nodes can be added under the order $p$ of the graph model. 

In this work, all Markov chain graphs are constructed to order $p = 1$ implying only single electron tunneling events. In future works, tunneling of multiple electrons, i.e., co-tunneling, can be incorporated by going to higher orders in the graph model.  

\begin{figure}
    \centering
    \includegraphics[width=\linewidth]{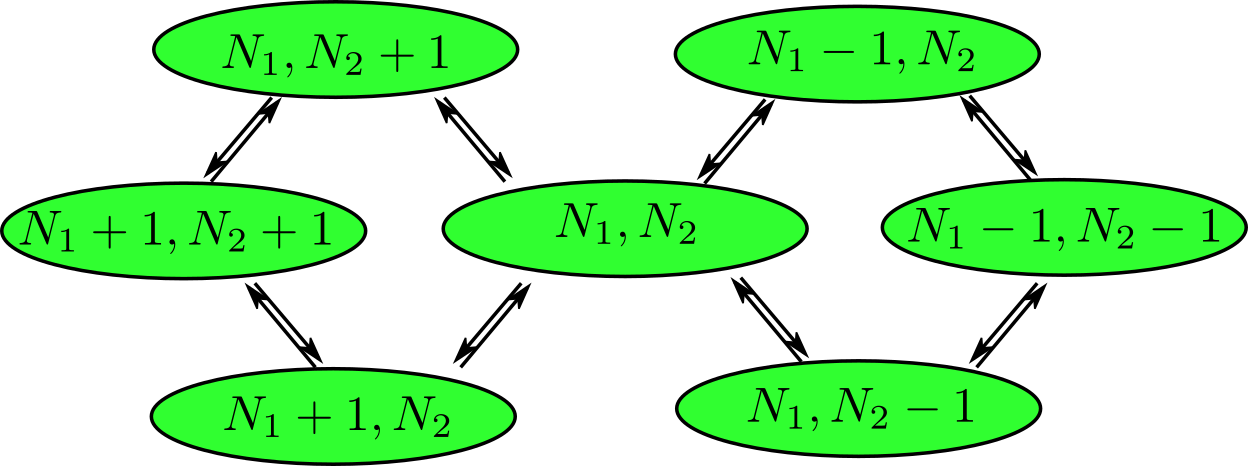}
\caption{An example of a Markov chain model for a double quantum dot}
\label{fig:markov}
\end{figure}

Fig.~\ref{fig:markov} shows an example of a simple state diagram for Markov chain in case of a double dot, with arrows representing the possible state transitions.

\subsection{Calculation of edge weights}
For two adjacent edges, the rate of going from one node to another is modeled as a product of two factors: a selection rule set by the capacitance model energies and a WKB tunnel rate. The rate for going from node 1 to 2 (arb. units) is given as:
\begin{equation}
R_{1 \rightarrow 2} = f_T(E_2 - E_1) \frac{p_{WKB}}{\tau}   
\end{equation}
where $f_T(E)=\frac{1}{1+\exp{{\frac{E}{kT}}}}$ is the Fermi function at temperature $T$, $E_2$ and $E_1$ are the capacitance model energies calculated for the charge configurations of nodes 2 and 1 respectively, $p_{WKB}$ is the WKB tunnel probability and $\tau = \frac{l_{dot}}{v_e}$, is the classical travel timescale inside a dot for an electron ($\l_{dot}$ is the dot size and $v_e$ is the classical electron velocity). 

The WKB tunnel probability $p_{WKB}$ is calculated by treating the electron as a free particle of energy equal to the Fermi level $\mu_F$ moving in a effective potential including inter-electron repulsion $V_{e}(x) = V(x) + \int K(x,x') n(x') dx'$. The tunneling probability is then computed as follows:

\begin{equation}
p_{WKB} = \exp{\left( -\int \frac{\sqrt{2 (V_{e}(x) - \mu)}}{\hbar} dx \right)}
\end{equation}
where the range of integration extends over the barrier region adjacent to the two locations through which the electron travels. The classical travel time scale, $\tau$, is calculated by treating the electron as a non-relativistic particle with kinetic energy $\mu_F$ and calculating the time it would take for the electron to transverse the extent of each island. 

\subsection{Current calculation}
Once the graph has been constructed and all edges have been assigned their requisite weights, calculation of the current proceeds by calculation of a stationary state of the Markov chain. Current $I$ (in arbitrary units) is given as: 
\begin{equation}
I = \sum_{(u,v), u \neq v} R_{u \rightarrow v} p(u) - R_{v \rightarrow u} p(v) 
\end{equation}
where the sum is over sets of nodes $(u,v)$ such that the transition from $u$ to $v$ corresponds to an electron transfer in a particular fixed direction (say left to right in the device) and $p(u)$ is the stationary probability of being on a node $u$.

Calculation of the stationary state is equivalent to finding the nullspace of the Markov matrix of the graph. Numerically, we calculated it using singular value decomposition (SVD) of the matrix by means of the LAPACK routine \texttt{\_gesdd} from \texttt{numpy.linalg} library. 

\section{Gate model}
\label{appendix:gate}
We assumed the top gates as cylindrical conductors of radius $r_0$ kept at a height $h$ from the electrons. This leads to a potential varying logarithmically with the distance from the gate.
A set of 4 parameters define each gate $(V_0,x_0,r_0,h)$. The profile for each gate is given as:

\begin{equation}
V(x) = \frac{V_0}{\log{h/r_0}} \log{\frac{\sqrt{(x-x_0)^2 + h^2}}{r_0}} \exp{-\frac{|x-x_0|}{\sigma_{sc}}}
\end{equation}
where $x_0$ defines the gate position, $V_0$ sets the height of the potential profile at $x = x_0$. $h$ controls the width of the profile. The term $e^{-\frac{|x-x_0|}{\sigma_{sc}}}$ has been added to take into account the screening due to the electron density present in the semiconductor. We used $\sigma_{sc} = 20\,$\si{\nano \meter} which is equal to the separation between adjacent gates.

\section{Device Parameters}
The following mean set of parameters was used for generating the datasets described in this work (for detailed description of the symbols see the appendix~\ref{appendix:tf} and appendix~\ref{appendix:gate}). They were randomly sampled from a Gaussian distribution with the mean values listed in the table. The standard deviation was set to 0.05 times the mean value. The idea behind generating the dataset in this fashion is to be able to train on all possible kinds of devices that one might expect in a lab as well as have the learning and tuning procedures robust against differences and imperfections across different devices.
\label{appendix:dataset}

For both set of devices, we used the common physical parameters, $K_0 = 10\,$\si{\milli \electronvolt} (sets the strength of the Coulomb interaction), 
$\sigma = 2\,$\si{\nano \meter} (prevents blowup at $x=0$ in the interaction), $g_0 = 0.5\,$\si{\electronvolt}$^{-1}\,$\si{\nano \meter}$^{-1}$ (sets the scale for the density of states) and $c_k = 1\,$\si{\milli \electronvolt \nano \meter} (kinetic term for the 2DEG). These values were chosen so that the quantum dots were in the few electron (1 to 10 electrons) regime. 

\subsection{3-gate device dataset}
The current and charge were calculated as a function of the plunger gate voltage, $V_P$, while the barrier gates, ($V_{B1}$ and $V_{B2}$) were held fixed. 
The total extent of the device was $(-40,40)\,$\si{\nano \meter}.
\begin{center}
\begin{tabular}{|c|c|c|c|c|}
    \hline
    Gate & $V_0$ (\si{\milli \volt}) & $x_0$ (\si{\nano \meter}) & $h$ (\si{\nano \meter}) & $r_0$ (\si{\nano \meter}) \\ \hline
    $b1$ & -200  & $-20$ & 50 & 5  \\  \hline
    $p$ & (0,400)  & 0 & 50 & 5\\  \hline
    $b2$ & -200 & 20 & 50 & 5\\  \hline
\end{tabular}
\end{center}

\subsection{5-gate device dataset}
The current and charge were calculated as a function of  $(V_{P1},V_{P2})$.
The total extent of the device was $(-60,60)$ \si{\nano \meter}.
\begin{center}
\begin{tabular}{|c|c|c|c|c|}
    \hline
    Gate & $V_0$ (\si{\milli \volt}) & $x_0$ (\si{\nano \meter}) & $h$ (\si{\nano \meter}) & $r_0$ (\si{\nano \meter})  \\ \hline
    $b1$ & -200  & $-40$ & 50  & 5\\  \hline
    $p1$ & (0,400)  & $-20$ & 50 & 5 \\  \hline
    $b2$ & -200 & 0 & 50 & 5\\  \hline
    $p2$ & (0,400) & 20 & 50 & 5\\  \hline
    $b3$ & -200 & 40 & 50 & 5\\  \hline
\end{tabular}
\end{center}

\section{Computing environment and TensorFlow parameters}\label{appendix:tensorflow}

We used TensorFlow, a machine learning API developed by Google to build and train the neural networks described in this work~\cite{tensorflow2015-whitepaper}. \texttt{tf.estimator} and \texttt{tf.layers} modules were used to create the deep and convolutional networks respectively.

The authors acknowledge the University of Maryland supercomputing resources (\url{http://hpcc.umd.edu}) made available for conducting the research reported in this paper.

All machine learning computations were performed on a 2015 Macbook Air laptop with 1.8GHz dual-core Intel Core i5 processor and 4 GB of RAM. No explicit GPU was used in training the neural networks, though in future studies with higher dimensions, it might be a necessity.

\end{document}